\documentclass[aps,12pt,preprint,groupedaddress,noshowpacs]{revtex4}

\usepackage{epsfig}
\usepackage{graphics}
\usepackage{slashed}
\usepackage{color}
\usepackage[citecolor=true]{hyperref}
\usepackage{multirow}
\usepackage{amsmath}

\begin{document}
%
\title{Pair production of heavy quarkonia  in the color evaporation model}
\author{\firstname{A.A.}\surname{ Chernyshev}} \email{aachernyshoff@gmail.com}
\affiliation{Samara National Research University,  Samara, 443086,
Russia}

\author{\firstname{V.A.}\surname{ Saleev}} \email{saleev@samsu.ru}
\affiliation{Samara National Research University, Samara, 443086,
Russia} \affiliation{Joint Institute for Nuclear Research, Dubna,
141980 Russia}

%
%
\begin{abstract}
In the article, we study  pair production of heavy quarkonia
$(J/\psi J/\psi, \Upsilon\Upsilon, \Upsilon J/\psi)$ in the improved
color evaporation model via the parton Reggeization approach. The
last one is based on high-energy factorization of hard processes in
multi-Regge kinematics, the Kimber-Martin-Ryskin-Watt model for
unintegrated parton distribution functions, and the Lipatov
effective field theory of Reggezied gluons and quarks. We compare
contributions from the single and double parton scattering
mechanisms in the pair production of heavy quarkonia. The numerical
calculations are performed with the Monte-Carlo event generator
KaTie. \vspace{0.2cm}
\end{abstract}
\maketitle
\section{Introduction}
Nowadays, pair production of heavy quarkonia in proton-proton
collisions at the LHC is studding intensively in the processes
\cite{CMS_2Psi,ATLAS_2Psi,LHCb_2Psi,CMS2Ups,D0PsiUps}
\begin{eqnarray}
&& p+p \rightarrow J/\psi + J/\psi +X,\label{PsiPsi} \\
&& p+p   \rightarrow \Upsilon + \Upsilon +X \label{UpsUps} ,\\
&& p+p \rightarrow\Upsilon + J/\psi +X \label{UpsPsi}.
\end{eqnarray}
These processes are very interesting for the theoretical study by
the following reasons. First, the production of four heavy quarks is
very good test for a perturbative quantum chromodynamics
(QCD)~\cite{pQCD}. The second ones, such processes of associated
production are strongly depended on the used model of heavy quark
hadronization into the heavy quarkonium, even more than in case of
inclusive heavy quarkonium production~\cite{CSM,NRQCD,ICEM,ICEM2}.
The last, but not the least reason is a possibility to study
directly double parton scattering (DPS) production
mechanism~\cite{DPS} which contribution to the pair production
cross-section estimates as more large in compare with the
conventional single parton scattering (SPS) scenario. To describe in
the collinear parton model (CPM) differential cross-section of the
pair heavy quarkonia with non-zero total transverse momenta, it is
needed to take into account next-to-leading order subprocesses with
at least five final particles, like this $g + g \to c + \bar c + b +
\bar b + g $. It is very complicated computing task in the pQCD and
CPM. Taking in mind the last sentence, we should use the high-energy
factorization (HEF) approach or $k_T$factorization
approach~\cite{kT1,kT2,kT3} which have used successfully to describe
a lot of data at the energy of the LHC. In the HEF, the initial
partons have non-zero transverse momenta and pair production of
heavy quarkonia may be described already in the leading order
approximation of pQCD in the subprocesses of gluon-gluon fusion via
SPS production mechanism
\begin{eqnarray}
&& R+R \rightarrow c+ \bar c+ c+ \bar c,  \label{cccc}\\
&& R+R   \rightarrow b+ \bar b+ b+ \bar b , \label{bbbb}  \\
 && R+R \rightarrow
c+ \bar c+ b+ \bar b. \label{ccbb}
\end{eqnarray}
or quark-antiquark annihilation
\begin{eqnarray}
&& Q_q+\bar Q_q \rightarrow c+ \bar c+ c+ \bar c, \\
&& Q_q+\bar Q_q   \rightarrow b+ \bar b+ b+ \bar b ,\\
&& Q_q+\bar Q_q \rightarrow c+ \bar c+ b+ \bar b.
\end{eqnarray}
 Here $R$ is the Reggeized gluon and $Q_q$ is the Reggeized quark (antiquark)
with four-momenta $q^\mu = x P^\mu+ q_T^\mu$, $P^\mu$ is the
relevant proton momentum, $q_T^\mu=(0, {\bf q_T}^\mu,0)$ and
$q^2=q_T^2=-{\bf q_T}^2$, $q=u,d,s$. We use HEF as it is formulated
in the parton Reggeization approach (PRA)~\cite{PRA1,PRA2,PRA3}.
Such approach has been used recently~\cite{PRA_cc} for the
description of the pair $J/\psi$ production cross section in the
improved color evaporation (ICEM) model~\cite{ICEM,ICEM2} with MC
event generator KaTie~\cite{katie}. At the presented paper, we study
associated production of $\Upsilon\Upsilon$ and $\Upsilon J/\psi$ in
the similar manner. The results for pair $J/\psi$ production are
also presented for completeness.

\section{Theoretical basics}
\subsection{Parton Reggeization approach}
\label{sec:PRA} The PRA is based on the HEF factorization justified
in the leading logarithmic approximation of the QCD at high
energies~\cite{kT1,kT2,kT3}. Dependent on transverse momentum,
unintegrated parton distribution functions (unPDF) of Reggeized
quarks and gluons are calculated in the model proposed earlier by
Kimber, Martin, Ryskin and Watt (KMRW)\cite{KMR,WMR}, but with our
sufficient modifications~\cite{PRA3}. Reggeized parton amplitudes
are constructed according to the Feynman rules of the L.N.~Lipatov
effective field theory (EFT) of Reggeized gluons and
quarks~\cite{Lipatov95, LipatovVyazovsky}. A review of the PRA can
be found in Refs.~\cite{PRA1,PRA2,PRA3}. Inclusion of high-order
corrections in the PRA  was studied in the
Refs.~\cite{NS2019,PRANLO1,PRANLO2,PRANLO3}.

In the PRA, the cross section of the inclusive process $p + p \to
{\cal Q} + X$ via SPS mechanism is related to the cross section of
the parton subprocess by the factorization formula
\begin{eqnarray}
  d\sigma^{\rm SPS} & = & \sum_{i, \bar{j}}
    \int\limits_{0}^{1} \frac{dx_1}{x_1} \int \frac{d^2{\bf q}_{T1}}{\pi}
{\Phi}_i(x_1,t_1,\mu^2)
    \int\limits_{0}^{1} \frac{dx_2}{x_2} \int \frac{d^2{\bf q}_{T2}}{\pi}
{\Phi}_{j}(x_2,t_2,\mu^2)\cdot d\hat{\sigma}_{\rm PRA},
\label{eqI:kT_fact}
\end{eqnarray}
where $t_{1,2} = - {\bf{q}}_{T 1,2}^2$, the cross section of the
subprocess with Reggeized partons $\hat{\sigma}_{\mathrm{PRA}}$ is
expressed in terms of squared Reggeized amplitudes
$\overline{|{\mathcal{A}}_{\mathrm{PRA}}|^2}$. The PRA
hard-scattering amplitudes are gauge invariant because the
initial-state off-shell partons are considered as Reggeized partons
of the gauge-invariant EFT for QCD processes in the MRK limit
~\cite{Lipatov95,LipatovVyazovsky}. The unPDFs in the modified KMRW
model are calculated by the formula~\cite{PRA3}
\begin{equation}
\Phi_i(x,t,\mu) = \frac{\alpha_s(\mu)}{2\pi}
\frac{T_i(t,\mu^2,x)}{t} \sum\limits_{j =
q,\bar{q},g}\int\limits_{x}^{1} dz\ P_{ij}(z) {F}_j \left(
\frac{x}{z}, t \right) \theta\left( \Delta(t,\mu)-z
\right),\label{uPDF}
\end{equation}
where $F_i(x,\mu_F^2) = x f_j(x,\mu_F^2)$. Here and below,
factorization and renormalization scales are equal, $\mu_F = \mu_R =
\mu$, and $\Delta(t,\mu^2)=\sqrt{t}/(\sqrt{\mu^2}+\sqrt{t})$ is the
KMRW-cutoff function~\cite{KMR}.  The modified unPDF
${\Phi}_i(x,t,\mu)$ should be satisfied exact normalization
condition:
\begin{equation}
\int\limits_0^{\mu^2} dt \Phi_i(x,t,\mu^2) =
{F}_i(x,\mu^2),\label{eq:norm}
\end{equation}
or
\begin{equation}
\Phi_i(x,t,\mu^2) = \frac{d}{dt}\left[ T_i(t,\mu^2,x){F}_i(x,t)
\right],\label{eq:sudakov}
\end{equation}
where $T_i(t,\mu^2,x)$ is the Sudakov form--factor, $T_i(t =
0,\mu^2,x) = 0$ and $T_i(t = \mu^2,\mu^2,x) = 1$. The explicit form
of the Sudakov form factor in the ({\ref{eq:sudakov}) was first
obtained in~\cite{PRA3}:
\begin{equation}
T_i(t,\mu^2,x) = \exp\left[-\int\limits_t^{\mu^2} \frac{dt'}{t'}
\frac{\alpha_s(t')}{2\pi} \left( \tau_i(t',\mu^2) + \Delta\tau_i
(t',\mu^2,x) \right) \right],\label{eq:sud}
\end{equation}
where
\begin{eqnarray*}
\tau_i(t,\mu^2) & = & \sum\limits_j \int\limits_0^1 dz\
zP_{ji}(z)\theta
(\Delta(t,\mu^2)-z), \label{eq:tau} \\
\Delta\tau_i(t,\mu^2,x) & = & \sum\limits_j \int\limits_0^1 dz\
\theta(z - \Delta(t,\mu^2)) \left[ zP_{ji}(z) -
\frac{{F}_j\left(\frac{x}{z},t \right)}{{F}_i(x,t)} P_{ij}(z)
\theta(z - x) \right].
\end{eqnarray*}
In contrast to the KMRW model, the Sudakov form factor
(\ref{eq:sud}) depends on $x$, which is necessary to preserve the
exact normalization (\ref{eq:norm}) for any $x$ and $\mu$. The gauge
invariance of amplitudes with Reggeized partons  in the PRA
guaranteed allows you to study any processes described non-Abelian
QCD structures. PRA has been successfully used for descriptions of
angular correlations in two-jet events~\cite{PRA1}, production of
the charm~\cite{Maciula:2016wci,Karpishkov:2014epa} and beauty
mesons~\cite{PRA2,Karpishkov:2016hnx}, charmonium in the NRQCD
~\cite{Saleev:2012hi,He:2019qqr}.

\subsection{Improved color evaporation model}
\label{sec:icem} The relevant description of the ICEM can be found
in the Ref.~\cite{ICEM_Ma_Vogt}. In the PRA, the $p_T$-spectra of
single $J/\psi(\Upsilon)$ is possible at the leading order
approximation of pQCD in the parton subprocesses
\begin{equation}
R + R \to c(b) + \bar{c}(\bar b) \label{RRcc}
\end{equation}
and
\begin{equation}
Q_q + \bar{Q}_q \to c(b) + \bar{c}(\bar b),\label{QQcc}
\end{equation}
where  $q = u, d, s$.

In the ICEM, the cross-section for the production of prompt
$J/\psi(\Upsilon)$-mesons is related to the cross-section for the
production of $c \bar{c}(b \bar b)$ pairs in the single parton
scattering (SPS) as follows:
\begin{eqnarray}
\sigma^{\rm SPS}(p + p \to J/\psi(\Upsilon) +
X)=\mathcal{F}^{\psi(\Upsilon)} \times
\int_{m_\psi(m_\Upsilon)}^{2m_D(2m_B)} \frac{d\sigma(p+p\to
c(b)+\bar c(\bar b)+X)}{dM}dM,\label{eq:ICEM1}
\end{eqnarray}
where $M$ is the invariant mass of the $c \bar{c}(b\bar b)$ pair
with 4--momentum $p_{c \bar{c}(b \bar b)}^\mu = p_{c(c)}^\mu +
p_{\bar{c}(\bar b))}^\mu$, $m_{\psi,\Upsilon}$ is the mass of the
$J/\psi(\Upsilon)$ meson and $m_D(m_B)$ is the mass of the lightest
$D(B)$ meson. To take into account the kinematical effect
associated with the difference between the masses of the
intermediate state and the final heavy quarkonium, the 4--momentum
of $c \bar{c}(b\bar b)$ pair and $J/\psi(\Upsilon)$ meson are
related by $p_{\psi(\Upsilon)}^\mu = (m_{\psi(\Upsilon)} / M) \,
p_{c \bar{c}(b \bar b)}^\mu$. The universal parameter
$\mathcal{F}^{\psi(\Upsilon)}$ is considered as a probability of
transformation of the $c\bar{c}(b \bar b)$ pair with invariant mass
$m_{\psi(\Upsilon)} < M < 2 m_{D(B)}$ into the prompt
$J/\psi(\Upsilon)$ meson.

The cross section for the production of a pair of prompt $J/\psi$
mesons ($\Upsilon\Upsilon$ or $\Upsilon J/\psi$) via the SPS is
related to the cross section for the production of two pairs
$c\bar{c}$ quarks in the following way
\begin{eqnarray}
&& \sigma^{\mathrm{SPS}}(p + p\to J/\psi + J/\psi + X) = \\
\nonumber &&=\mathcal{F}^{\psi \psi} \times \int_{m_\psi}^{2m_D}
\int_{m_\psi}^{2m_D} \frac{d\sigma(p + p \to c_1 + \bar{c}_1 + c_2 +
\bar{c}_2 + X)}{dM_1 dM_2} dM_1 dM_2,\label{eq:ICEM2}
\end{eqnarray}
where $M_{1, 2}$ are the invariant masses of $c \bar{c}$ pairs with
4--momenta $p_{c \bar{c}1}^\mu = p_{c1}^\mu + p_{\bar{c}1}^\mu$ and
$p_{c \bar{c}2}^\mu = p_{c2}^\mu + p_{\bar{c}2}^\mu$. Parameter
$\mathcal{F}^{\psi \psi}$ is the probability of transformation of
two pairs $c\bar c$ with invariant masses $m_\psi < M_{1 ,2} < 2
m_D$ into two $J/\psi$ mesons.

In the DPS approach~\cite{DPS}, the cross section for the production
of a $J/\psi$ pair is written in terms of the cross sections for the
production of single a $J/\psi$ in two independent subprocesses
\begin{equation}
\sigma^{\mathrm{DPS}}(p + p \to J/\psi + J/\psi + X)= \frac{
\sigma^{\mathrm{SPS}}(p + p \to J/\psi + X_1) \times
\sigma^{\mathrm{SPS}}(p + p \to J/\psi + X_2)} {N_{f}
\sigma_{\mathrm{eff}}} \label{eq:DPS},
\end{equation}
where $N_f=2$ for the $J/\psi J/\psi (\Upsilon\Upsilon)$ pair
production, $N_f=1$ for the $J/\psi \Upsilon$ pair production and
the parameter $\sigma_{\mathrm{eff}}$, which controls the
contribution of the DPS mechanism, is considered as free parameter.
Thus, at fitting  cross sections for the pair
$J/\psi(\Upsilon)$--meson production, we assume that the parameter
$\mathcal{F}^{\psi,\Upsilon}$ is fixed, and the parameters
$\mathcal{F}^{\psi \psi, \Upsilon \Upsilon}$ and
$\sigma_{\mathrm{eff}}$ are free parameters.

\section{Numerical calculations and results}

Recently, a new approach to obtaining gauge invariant amplitudes
with off-shell initial-state partons in hard subprocesses at high
energies has been proposed. The method is based on the use of spinor
amplitude formalism and recurrence relations of the BCFW
type~\cite{hameren1,hameren2}.  As it was demonstrated recently,
this formalism~\cite{hameren1,hameren2}, which is implemented in
event generator KaTie, for numerical amplitude generation is
equivalent to amplitudes built according to Feynman rules of the
Lipatov EFT at the level of tree diagrams~\cite{PRA1,PRA2,kutak}.
Such a way, at the stage of numerical calculations, we use the
Monte-Carlo event generator KaTie~\cite{katie} for calculating the
proton-proton cross sections, as well as it was done previously in
Ref.~\cite{PRA_cc}. The accuracy of numerical calculations for total
proton-proton cross sections is equal to 0.1\%.

\subsection{Associated $J/\psi J/\psi$ production}
The complete results for $J/\psi J/\psi$ pair production is
collected in our recent paper~\cite{PRA_cc}. Here we present only
main ones. We obtain a quite satisfactory description for the single
prompt $J/\psi$ $p_T-$spectra and cross sections in the ICEM using
the PRA at the wide range of the collision energy. The obtained
values of the hadronization parameter $\mathcal{F}^\psi$ depend on
energy, and such dependence can be approximated by the formula
$\mathcal{F}^\psi(\sqrt{s})=0.012+0.952 (\sqrt{s})^{-0.525}$, see
the Fig.~\ref{fig:1}. The data for the pair $J/\psi$ production
cross sections at the energy range $7-13$ TeV can be fitted
self-consistently with two free parameters
${\mathcal{F}}^{\psi\psi}$ and $\sigma_{\rm eff}$. We have found the
best fit with ${\mathcal{F}}^{\psi\psi}\simeq 0.02$ and $\sigma_{\rm
eff} \simeq 11.0$ mb, see the Fig.~\ref{fig:2}, were
$$
x = \sum_{k = 1}^n \frac{|\sigma^{\mathrm{exp}}_k -
\sigma^{\mathrm{theor}}_k|} {\Delta\sigma^{\mathrm{exp}}_k}
$$
and the sum is taken over all cross sections of three experiments:
CMS~\cite{CMS_2Psi}, ATLAS~\cite{ATLAS_2Psi} and
LHCb~\cite{LHCb_2Psi}. It is interesting that we find
${\mathcal{F}}^{\psi\psi} \approx {\mathcal{F}}^{\psi}$
\cite{PRA_cc}. Complete set of the plots for differential cross
sections in $J/\psi J/\psi$ pair production are presented in the
Ref.~\cite{PRA_cc}.

\subsection{Associated $\Upsilon\Upsilon$ production}
 As in the case of single prompt $J/\psi$ production, we obtain a quite satisfactory
description for the single prompt $\Upsilon(1S)$ total cross
sections in the ICEM using the PRA at the wide range of the
collision
energy~\cite{AFS,CDF,LHCb1,ATLAS,CMS,LHCb2,ALICE,LHCb3,LHCb4}. The
obtained values of the hadronization probability
$\mathcal{F}^\Upsilon$ depend on energy by the formula
$\mathcal{F}^\Upsilon(\sqrt{s})=0.012+4.166 (\sqrt{s})^{-0.677}$,
see the Fig.~\ref{fig:3}.

Pair $\Upsilon\Upsilon$ production cross-section was measured by the
CMS Collaboration at the energy $\sqrt{s}=13$ TeV~\cite{CMS2Ups2}.
Taking into account the contributions of the SPS and the DPS
production mechanisms, parameter $\mathcal{F}^{\Upsilon \Upsilon}$
is obtained by the fit of the rapidity difference spectrum and the
azimuthal angle difference spectrum, as it is shown in
Figs.~\ref{fig:4} and \ref{fig:5}. Here, the parameter
$\mathcal{F}^{\Upsilon}$ is fixed by the fitting of single
$\Upsilon$ production and the universal parameter is taken
$\sigma_{\mathrm{eff}}=11.0\pm 0.2$  mb, as it was obtained by the
fitting of single and pair $J/\psi$ production cross-sections.  We
find that $\mathcal{F}^{\Upsilon \Upsilon}\simeq
\mathcal{F}^{\Upsilon} \simeq 0.05$, the same as it is for pair
$J/\psi$ production. Such a way, as it is shown in Figs.~\ref{fig:4}
and \ref{fig:5}, SPS contribution in the pair $\Upsilon$ production
cross-section is about 10 \% only and DPS contribution dominates.

\subsection{Associated $\Upsilon J/\psi$ production}
The $\Upsilon J/\psi$ pair production cross section and the
azimuthal angle difference spectrum were measured by D0
Collaboration at the energy $\sqrt{s}=1.96$ TeV~\cite{D0PsiUps}.
When the parameters $\sigma_{\mathrm{eff}}=11$ mb,
$\mathcal{F}^\Upsilon=0.039$, and $\mathcal{F}^{\psi}=0.044$, have
been fixed, we predict the hadronization probability
$\mathcal{F}^{\psi \Upsilon}$ approximately equal to
{$\mathcal{F}^{\psi} \times \mathcal{F}^{\Upsilon} = 0.002$}. At
this values of the hadronization probabilities, the SPS contribution
in $J/\psi \Upsilon$ pair production is negligibly small, see
Fig.~\ref{fig:6}.

\section{Conclusions}

We obtain a quite satisfactory description for the single prompt
$J/\psi$ and $\Upsilon$ production $p_T-$spectra and cross sections
in the ICEM using the PRA at the wide range of the collision energy.
It is obtained that the values of the hadronization probability
$\mathcal{F}^{\psi,\Upsilon}$ are depended on energy.

Both mechanisms, SPS and DPS, for the pair $J/\psi$ and $\Upsilon$
pair production have been considered. The data for the pair $J/\psi$
and $\Upsilon$ production cross sections at the energy range $7-13$
TeV can be fitted self-consistently with free parameters
${\mathcal{F}}^{\psi\psi}$, ${\mathcal{F}}^{\Upsilon\Upsilon}$ and
$\sigma_{\rm eff}$. We have found the best fit with $\sigma_{\rm
eff} \simeq 11.0\pm 0.2$ mb, which is in a good agreement with
another theoretical estimates. We find the dominant role of the DPS
mechanism  in $\Upsilon\Upsilon$ and $J/\psi \Upsilon$ pair
production. In case of pair $J/\psi$ production, the DPS mechanism
absolutely dominates only at the forward region of $J/\psi$
rapidities.

The important phenomenological finding is that
${\mathcal{F}}^{\psi\psi}\simeq {\mathcal{F}}^{\psi} \neq
{\mathcal{F}}^{\psi}\times {\mathcal{F}}^{\psi}$ and
${\mathcal{F}}^{\Upsilon\Upsilon}\simeq {\mathcal{F}}^{\Upsilon}\neq
{\mathcal{F}}^{\Upsilon}\times {\mathcal{F}}^{\Upsilon}$, but
${\mathcal{F}}^{\Upsilon \psi}\simeq {\mathcal{F}}^{\Upsilon}\times
{\mathcal{F}}^{\psi}$ . This result demonstrate the important
property of quantum identity and Pauli principle for quarks.

\section*{Acknowledgments}
We are grateful to A. Van Hameren for advice on the program KaTie.
The work was supported by the Ministry of Science and Higher
Education of the Russian Federation, project FSSS-2020-0014.

%
\clearpage
%
%
%
%
%
\newpage
%

\begin{figure}
\begin{center}
\includegraphics[width=0.75\textwidth,angle=0]{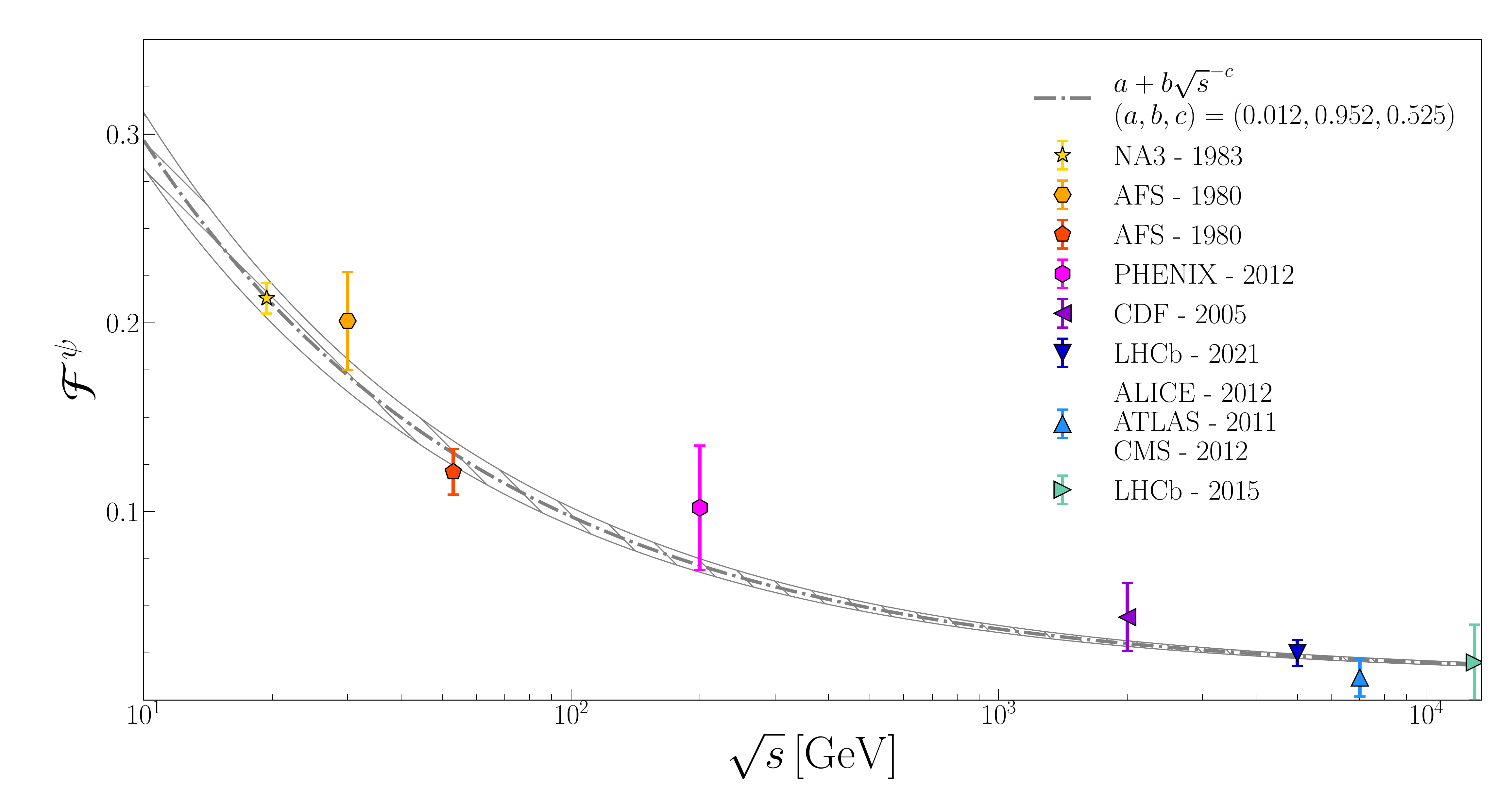}
    \end{center}
\caption{ The hadronization probability $\mathcal{F}^{\psi}$ as a
function of proton collision energy $\sqrt{s}$. The corridor between
the upper and lower lines demonstrates the uncertainty from the hard
scale variation by the factor $\xi = 2$  and the $c$-quark mass from
$1.2$ to $1.4$ GeV.}\label{fig:1}
\end{figure}
\newpage
%


\newpage


\begin{figure}
\begin{center}
\includegraphics[width=0.75\textwidth,angle=0]{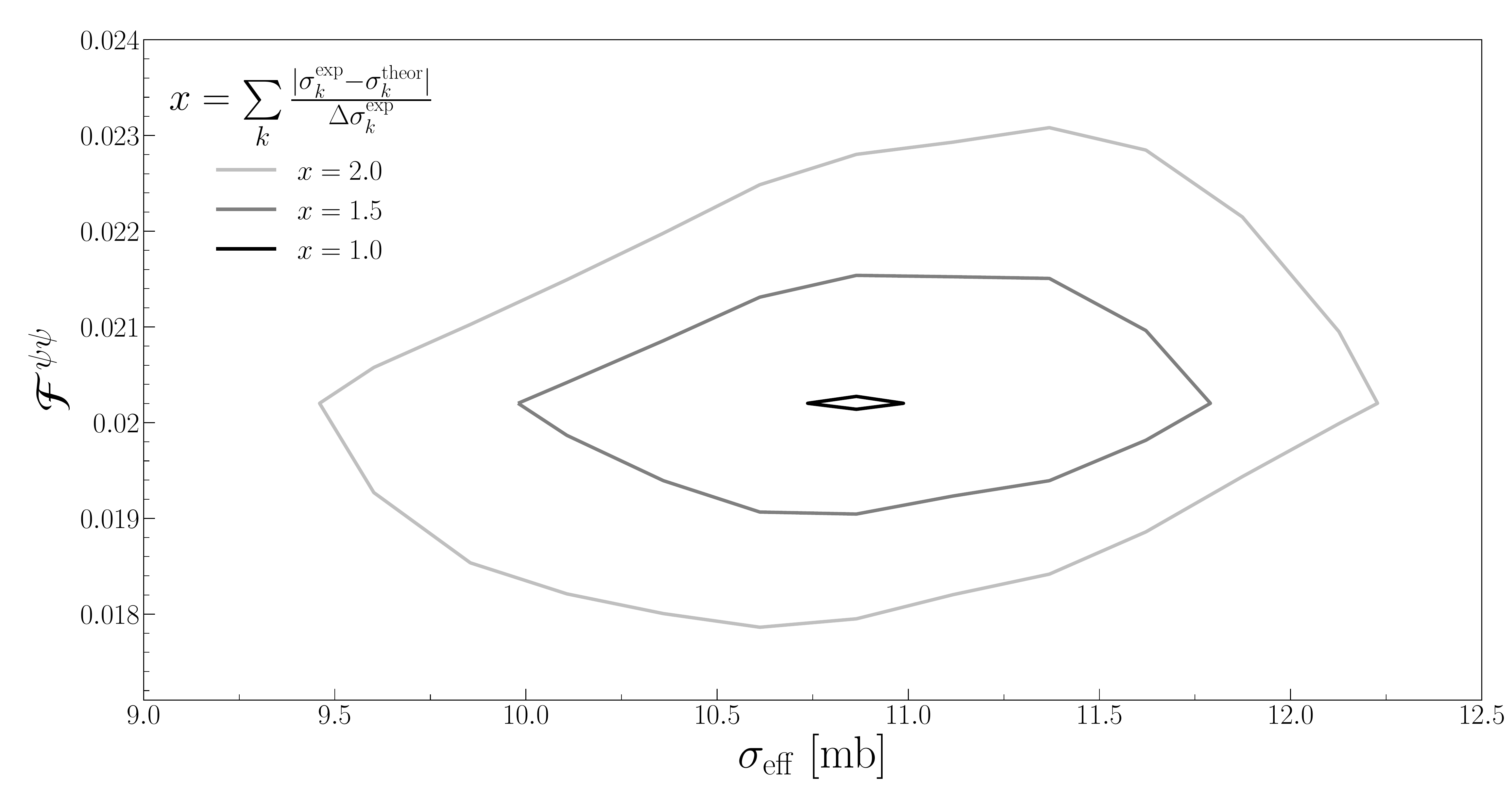}
    \end{center}
\caption{ Regions of the probabilities $\mathcal{F}^{\psi \psi}$ and
$\sigma_{\mathrm{eff}}$ in the ICEM for pair $J/\psi$ production,
obtained as a result of data fitting. Isolines correspond to $x =
1.0, 1.5$ and $2.0$. }\label{fig:2}
\end{figure}


\begin{figure}
\begin{center}
\includegraphics[width=0.75\textwidth,angle=0]{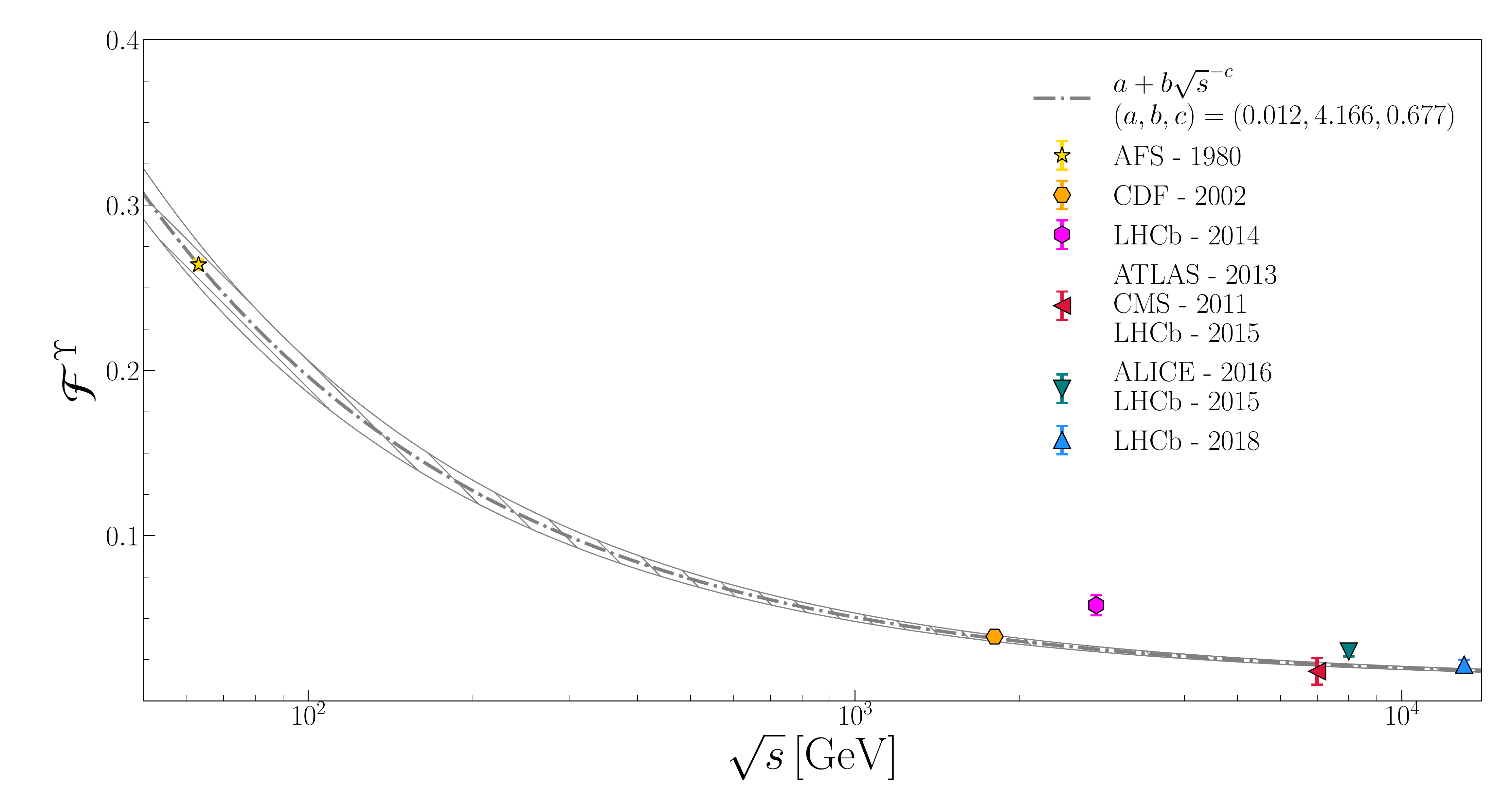}
    \end{center}
\caption{ The hadronization probability $\mathcal{F}^{\Upsilon}$ as
a function of proton collision energy $\sqrt{s}$. The corridor
between the upper and lower lines demonstrates the uncertainty from
the hard scale variation by the factor $\xi = 2$  and the $b$-quark
mass from $4.5$ to $4.75$ GeV.}\label{fig:3}
\end{figure}

\newpage


%


\begin{figure}
\begin{center}
\includegraphics[width=0.75\textwidth,angle=0]{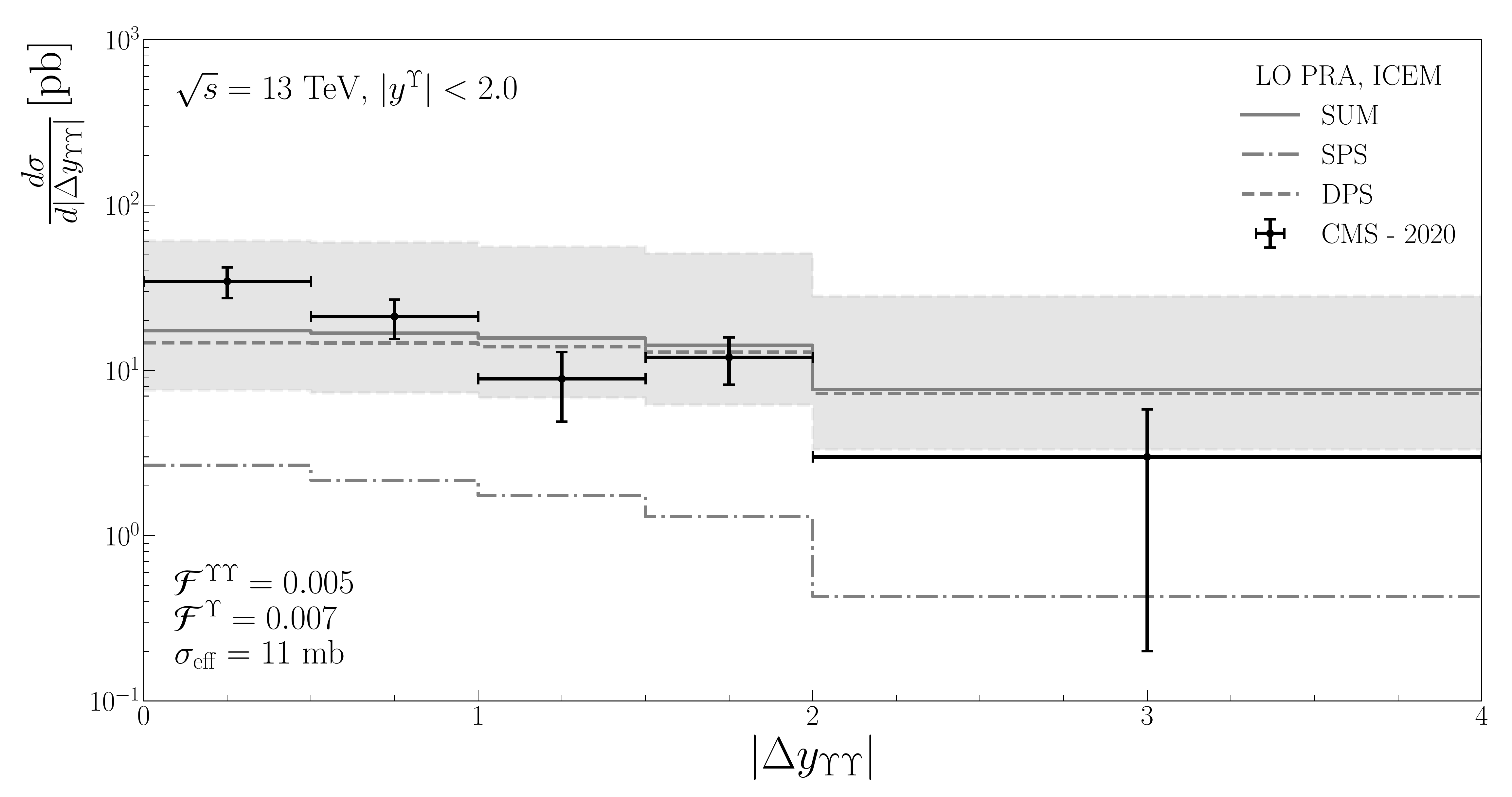}
    \end{center}
\caption{ Different spectra of pair $\Upsilon$ production as a
function of rapidity difference $|\Delta y_{\Upsilon\Upsilon}|$. The
data are from CMS collaboration at the $\sqrt{s}=13$ TeV
\cite{CMS2Ups2}.}\label{fig:4}
\end{figure}


\begin{figure}
\begin{center}
\includegraphics[width=0.75\textwidth,angle=0]{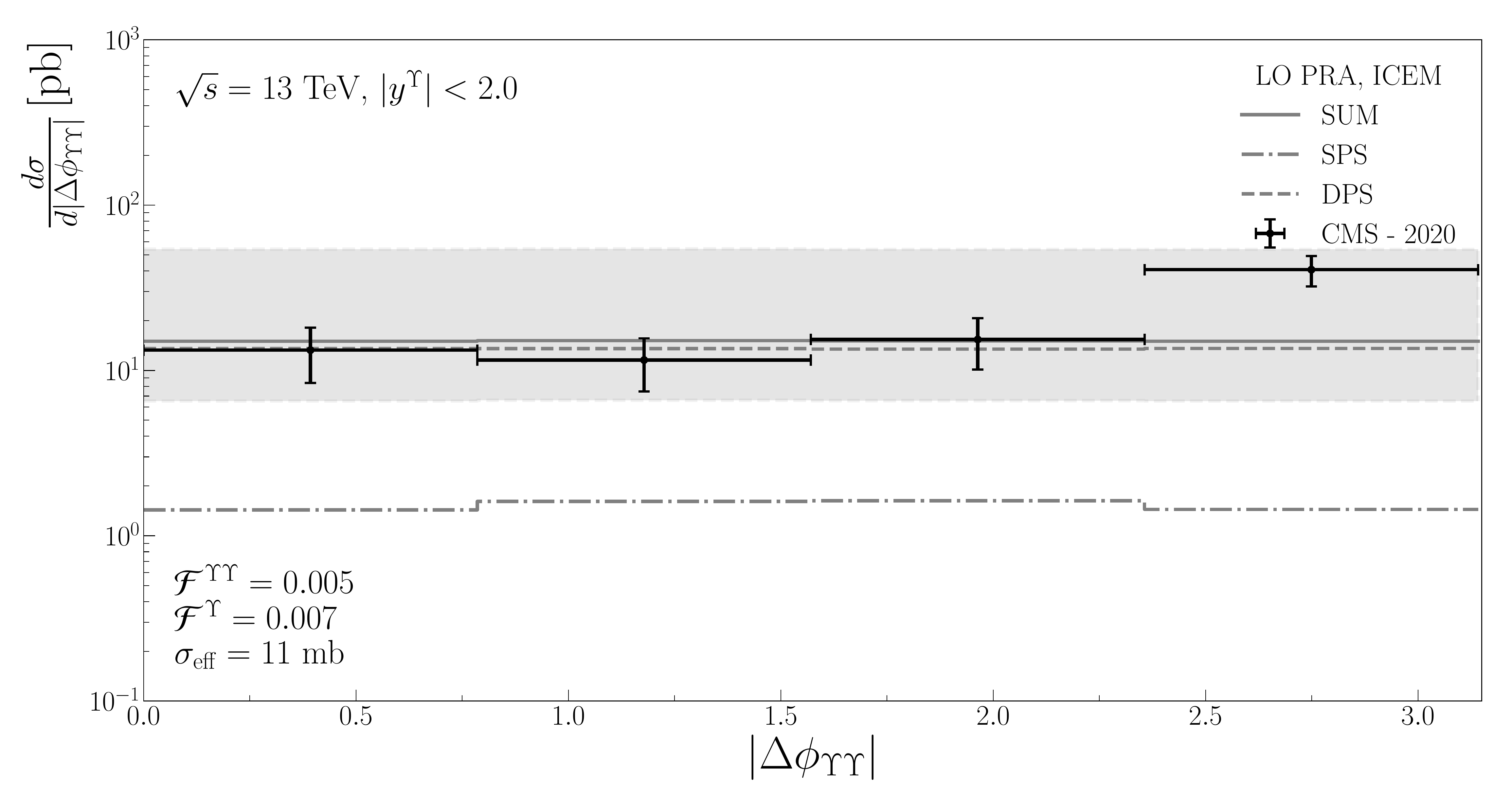}
    \end{center}
\caption{  Different spectra of pair $\Upsilon$ production as a
function of azimuthal angle difference $|\Delta
\phi_{\Upsilon\Upsilon}|$. The data are from CMS collaboration at
the $\sqrt{s}=13$ TeV \cite{CMS2Ups2}.}\label{fig:5}
\end{figure}


\begin{figure}
\begin{center}
\includegraphics[width=0.75\textwidth,angle=0]{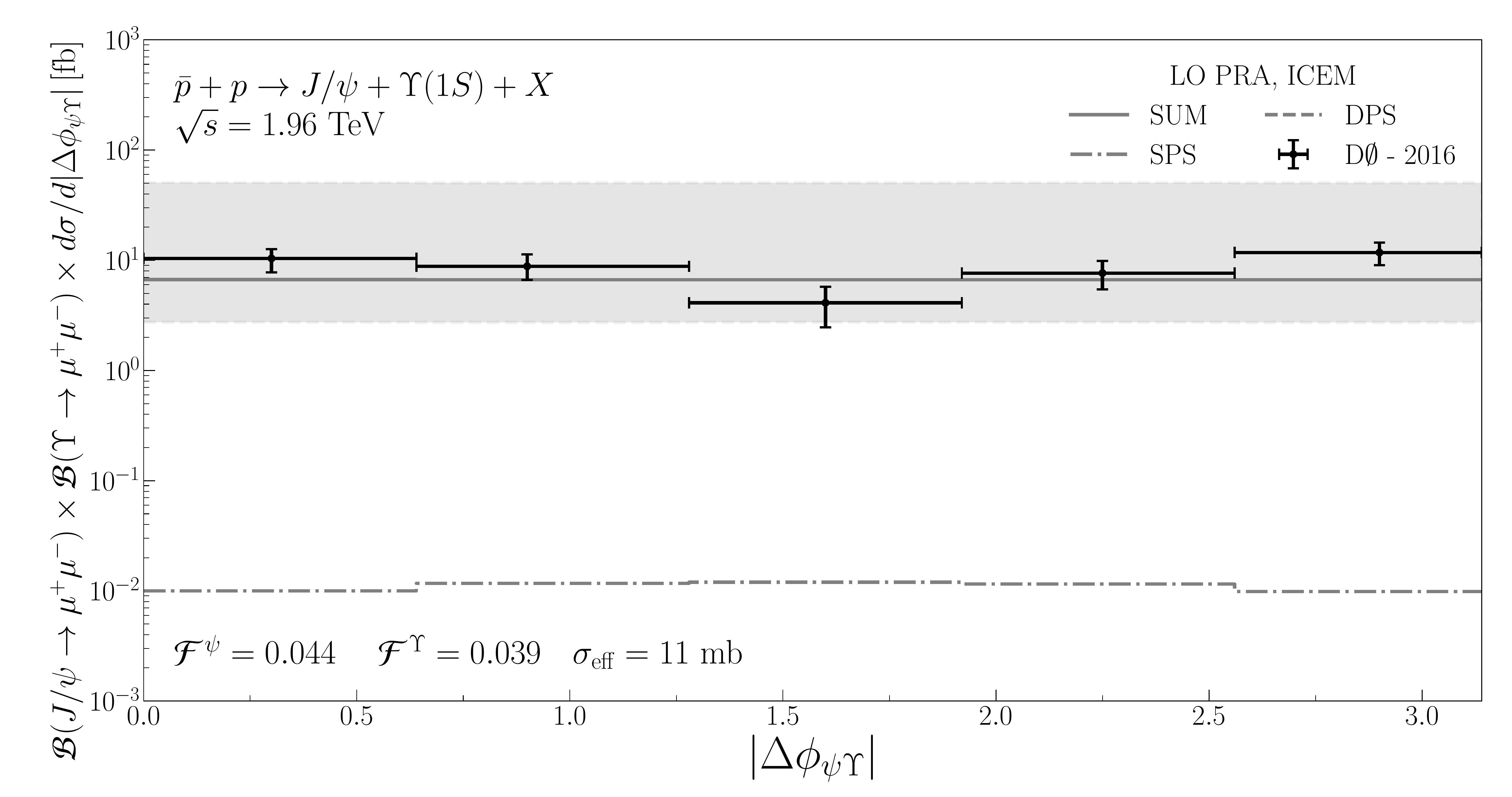}
    \end{center}
\caption{  Different spectra of pair $J/\psi \Upsilon$ production as
a function of azimuthal angle difference $|\Delta \phi_{J/\psi
\Upsilon}|$. The data are from D0 collaboration at the
$\sqrt{s}=1.96$ TeV \cite{D0PsiUps}.}\label{fig:6}
\end{figure}

\clearpage
\bibliography{bibliography_mifi_2}

\end{document}